\documentclass[12pt]{article}
 \usepackage{helvet}
\usepackage{amsmath,amsthm,amssymb,setspace,enumitem,epsfig,rotating,titlesec,verbatim,color,array}

\usepackage[round,colon]{natbib}
\usepackage{eurosym}
\usepackage{lineno}
\usepackage{textcomp}
\usepackage{todonotes}
\usepackage{graphicx}
\usepackage{hyperref}
\hypersetup{
  colorlinks=true,       
    linkcolor=blue,          
    citecolor=darkgreen,        
    filecolor=magenta,      
    urlcolor= black           
}
\usepackage{color}
	 \definecolor{darkgreen}{rgb}{0,0.5,0}
	 \definecolor{magenta}{rgb}{0,0,0.75}
	 
\usepackage{paralist}
\usepackage[top=2cm,left=2cm,right=2cm,bottom=2cm]{geometry}
\usepackage{setspace}
\doublespace

\usepackage{arydshln,booktabs}
\newcolumntype{C}[1]{>{\centering\arraybackslash}p{#1}} 
\begin{document}

\date{}
\title{Chaotic provinces in the kingdom of the Red Queen}

\author{Hanna Schenk$^{1}$, Arne Traulsen$^{1}$ and Chaitanya S. Gokhale$^{1,2,\ast}$ \\
\vspace{-2mm}\normalsize $^{1}$Department of Evolutionary Theory, Max Planck Institute for Evolutionary Biology, \\
\vspace{-2mm}\normalsize August-Thienemann Str-2, 24306 Pl\"on, Germany\\
\vspace{-2mm}\normalsize $^{2}$New Zealand Institute for Advanced Study, Massey University, Albany, Private Bag 102904\\
\vspace{-2mm}\normalsize North Shore Mail Centre, 0745, Auckland, New Zealand\\
\vspace{-2mm}\normalsize $^{\ast}$gokhale@evolbio.mpg.de} 
\date{\normalsize\today}

\maketitle

\noindent
{\bf Running title:} Chaotic Red Queen
\\

\newpage


\section*{Abstract}
The interplay between parasites and their hosts is found in all kinds of species and plays an important role in understanding the principles of evolution and coevolution. 
Usually, the different genotypes of hosts and parasites oscillate in their abundances.
The well-established theory of oscillatory Red Queen dynamics proposes an ongoing change in frequencies of the different types within each species.
So far, it is unclear under what conditions Red Queen dynamics persists, 
especially when the number of types per species increases. 
Some models show that with many types of hosts and parasites or more species chaotic dynamics occur.
In our analysis, an arbitrary number of types within two species are examined in a deterministic framework with constant or changing population size and very simple interactions.
This general framework allows for analytical solutions for internal fixed points and their stability.
The numerical analysis shows that for two species, once more than two types are considered per species, irregular dynamics in their frequencies can be observed in the long run.
The nature of the dynamics depends strongly on the initial configuration of the system; 
the usual regular Red Queen oscillations are only observed when all types initially have similar abundance. 

\section{Introduction}

Studying host-parasite coevolution using mathematical models has led to substantial advances in our understanding of the dynamics of the interaction.
For example, hypothesising the role of reciprocal selection between the antagonistic species in the evolution of virulence and tolerance.
We specifically focus on the Red Queen hypothesis \citep{vanValen:EvoTheo:1973,stenseth:Evolution:1984,dieckmann:JTB:1995,clay:ARPh:1996,salathe:TREE:2008}. 
This hypothesis has been used in a broad context, leading to multiple definitions \citep{brockhurst:PRSB:2014,rabajante:SciRep:2015}. 
According to Van Valen, the maintenance of biodiversity is possible as long as the species displace each other, or when the resource distribution changes over time \citep{vanValen:EvoTheo:1973}. 
However, the different definitions are underlined by the presence of the typical dynamics expected within a species, namely Red Queen oscillations.
These oscillations imply an interaction where the increase in the relative abundance of a certain type within a species indicates an equal decrease in relative abundance of another type \citep{maynard-smith:AmNat:1976b,VanValen:AmNat:1977}.
In the context of hosts and parasites, indications for such oscillations in densities have been empirically confirmed, for example, in dormant stages of the water flea \textit{Daphnia magna} from pond sediments \citep{decaestecker:Nature:2007} and freshwater snails \textit{Potamopyrgus antipodarum} \citep{koskella:Evolution:2009}. 
Another experiment \citep{yoshida:Nature:2003} on evolution of prey also showed that more variability lead to different phase lags in oscillating predator-prey frequencies. 
However, while it is already difficult to analyse these dynamics experimentally over a single cycle, analysing the long term dynamics of such systems is challenging. 

The co-evolution of hosts and parasites has for example been used to explain sexual reproduction \citep{lively:JH:2010}. 
However, when multilocus genetics is at play, features of co-evolution models, such as the maintenance of polymorphism and evolution of sex, depend on the exact interaction patterns \citep{frank:EEcol:1993,parker:EEcol:1996,frank:EEcol:1996,sasaki:PRSB:2000,metzger:Evolution:2016}. 
Exploring a variety of interaction patterns between multiple types of hosts and parasites, we show that short-term oscillations as the ones observed experimentally can be recovered in virtually all of these models, but one has to be very careful in extrapolating this kind of dynamics over a wider time horizon.

Mathematical models of host-parasite interactions with two types have been extensively analysed.
A specific experiment in \textit{Daphnia magna} \citep{carius:Evolution:2001} showed considerable variation in hosts (susceptibility) and parasites (infectivity) of nine distinct types, which illustrates the necessity of considering models with more than two types.
A recent model \citep{rabajante:SciRep:2015} based on ordinary differential equations numerically explored increasing numbers of types.
The result strengthened the theory of the oscillatory Red Queen dynamics. 
In such numerical models, a broader exploration of the parameter space can lead to more general results and show the robustness of models.
Here, we take a different approach and ask: How complicated can a model become before the regular frequency dependent oscillations are lost?
To tackle this question, we used analytical tools in addition to numerical integration.

Another aspect to be considered is the impact of population size \citep{papkou:Zoology:2016}:
A host population suffering from
intense parasite pressure should decrease in absolute size. 
Similarly, a parasite population not finding sufficient hosts should decrease in absolute size.
Recently, the impact of such changing population sizes was studied for two types \citep{gokhale:BMCEvolBio:2013,song:BMCEvolBio:2015}, 
including matching alleles \citep{frank:PRSB:1993} or the gene-for-gene type of interactions \citep{flor:PHYTO:1955,engelstaedter:AMNAT:2015, agrawal:EER:2002}.
By adjusting the birth rates of hosts and death rates of parasites to include frequency dependence, we can impose a constant population size.
Such a transformation makes the underlying model identical to the replicator dynamics \citep{taylor:MB:1978,hofbauer:book:1998,schuster:JTB:1983}.

Here, we extend the two approaches with changing and constant population size to an arbitrary number of types of hosts and parasites.
As a simple example, we first consider the matching allele model: Each host can only be infected by its specific parasite type, and each parasite can affect only the matching host.
Next, cross-infectivity is incorporated so that two genetically similar parasites (neighbours to the focal parasite) can additionally infect a particular host in an equally robust manner and vice versa: each parasite can infect not only one host but also two more closely related hosts.
Finally, a generalised cross-infection model is briefly analysed. In this model, different infectivity magnitudes can be realised for each parasite type.

\section{Model}
\subsection{Interactions between hosts and parasites}
\label{sec:model}
The number of parasites affecting a focal host (and vice versa) and the strength of the interactions are key components for models of host-parasite coevolution. 
Three possible models are depicted in Table \ref{tab:interactions},
where fitness effects are collected in a matrix, which intuitively describes the influence of each type within one species on each type within the other species.
Assuming $n$ types of hosts and $n$ types of parasites, 
the $n \times n$ matrix $M^H$ describes the average loss of fitness hosts suffer from specific parasite types and the $n \times n$ matrix $M^P$ describes the average gain of fitness parasites extract from the interaction.
For example $\left(M^H\right)_{2,4}$ is the fitness loss that host 2 suffers from parasite type 4. 
On the other hand, $\left(M^P\right)_{4,2}$ is the fitness gain obtained by parasite 4 from host 2.\\

\begin{table}
	\centering
	\begin{tabular}{lcc}
		\toprule \addlinespace
		\multicolumn{1}{l}{Model}& \multicolumn{1}{c}{$M^P$}& \multicolumn{1}{c}{$M^H$} \\
		\cmidrule(lr){1-3} 
		\addlinespace
		\small{Matching alleles} & $\left( \begin{matrix}
		+1 & 0 & \cdots & 0 \\
		0 & +1 & \cdots & 0 \\
		\vdots  & \vdots  & \ddots & \vdots  \\
		0 & 0 & \cdots & +1
		\end{matrix} \right)$ & $\left( \begin{matrix}
		-1 & 0 & \cdots & 0 \\
		0 & -1 & \cdots & 0 \\
		\vdots  & \vdots  & \ddots & \vdots  \\
		0 & 0 & \cdots & -1
		\end{matrix} \right)$  \\
		\addlinespace[0.4cm]
		\small{Cross-infection} & $\left( \begin{matrix}
		+1 	&  +1		& 0  	& 0  	&\cdots	& +1 		\\
		+1 	&  +1 	&  +1 	& 0  	&\cdots	& 0 		\\
		0		&  +1 	&  +1 	&  +1 	&\cdots	& 0 		\\
		\vdots	&\vdots	&\vdots	&\vdots	&\ddots	& \vdots	\\
		0		& 0		& \cdots	&  +1		&  +1		&  +1		\\		 
		+1 		& 0 		& \cdots& 0		&  +1		&  +1		
		\end{matrix} \right)$ & $ \left( \begin{matrix}
		-1 	& -1 	& 0  	& 0  	&\cdots	& -1		\\
		-1 	& -1 	& -1 	& 0  	&\cdots	& 0 		\\
		0		& -1 	& -1 	& -1 	&\cdots	& 0 		\\
		\vdots	&\vdots	&\vdots	&\vdots	&\ddots	& \vdots	\\
		0		& 0		& \cdots	& -1		& -1		& -1		\\		 
		-1		& 0 		& \cdots& 0		& -1		& -1		\\
		\end{matrix} \right)$\\
		\addlinespace[0.4cm]
		\small{\begin{tabular}[x]{@{}c@{}}Generalised\\cross-infection\end{tabular}} & $ \left( \begin{matrix}
		\alpha_1	& \alpha_2 	& \cdots &\alpha_{n-1}& \alpha_n \\
		\alpha_n 	& \alpha_1 &\alpha_2	& \hdots & \alpha_{n-1} \\
		\vdots 	& \alpha_n  	& \alpha_1 & \hdots& \vdots  \\
		\alpha_3 & \vdots & \vdots & \ddots & \alpha_2\\
		\alpha_2 & \alpha_3	& \cdots &\alpha_n	&  \alpha_1
		\end{matrix} \right)$ & $\left( \begin{matrix}
		-c \alpha_1	& -c \alpha_n 	& \cdots &-c \alpha_{3}& -c \alpha_2 \\
		-c \alpha_2 	& -c \alpha_1 &-c \alpha_n	& \hdots & -c \alpha_{3} \\
		\vdots  	& -c \alpha_2  	& -c \alpha_1 & \hdots& \vdots  \\
		-c \alpha_{n-1} & \vdots & \vdots & \ddots & -c \alpha_n\\
		-c \alpha_n & -c \alpha_{n-1}	& \cdots &-c \alpha_2	&  -c \alpha_1
		\end{matrix} \right)$ \\
		\addlinespace \bottomrule 
	\end{tabular}
	\caption{\textbf{Interaction models:}  
	$M^P$ is the fitness gain of a parasite (row) achieved by exploiting a specific host (column).
	 $M^H$ is the fitness loss of a host (row) caused by a parasite (column).} \label{tab:interactions}
\end{table}

To introduce host-parasite dynamics, we focus on the matching allele model first \citep{grosberg:Science:2000,carius:Evolution:2001}, where only fixed pairs of hosts and parasites can directly interact with each other (Tab.~\ref{tab:interactions}, Matching alleles). 
Interactions with all other partners are neutral and do not influence fitness.
In a cross-infection model, it is instead assumed that neighbouring parasite types are genotypically or phenotypically similar in their infectivity (Tab.~\ref{tab:interactions}, Cross-infection).
This also applies to each host and its neighbours which have not developed resistance and are therefore susceptible to a specific parasite type which now benefits from three host types.
We assume that cross-infectivity follows periodic boundary conditions where types $1$ and $n$ can also interact with three types of the other species.
Finally, in our generalised cross-infection model, hosts have a positive effect $\alpha_i$ on parasites which have a negative effect $-c\alpha_i$ with $c>0$ on the hosts (Tab.~\ref{tab:interactions}, Generalised cross-infection). 
Every diagonal has a different value, which leads to $n$ interaction parameters.
This means that parasite $i$ has the same negative effect $-c\alpha_{k+1}$ on host $i+k$ as parasite $j$ on host $j+k$. 
In addition, we assume that host $i$ has the positive effect $\alpha_{k+1}$ on parasite $i-k$.
The restriction $M^H=-c \cdot \left(M^P\right)^T$ ensures a scaled effect of the interaction partners. 
In this way we can for example envision a scenario in which there are matching hosts and parasites (the main diagonal) and the effect they exert on each other declines with distance between them, 
$\alpha_1>\alpha_2>\alpha_3>\hdots$.

We stress that these matrices are not chosen to represent a particular biological system. 
Instead, our approach is to consider more complex models beyond the matching allele models and to analyse their dynamics. 
The fitness effects represented in the matrices are included in the models.
In the following, we differentiate between models with changing population size (Lotka--Volterra equations) and models with constant population size (replicator dynamics).

\subsection{Changing population size}

The classical Lotka--Volterra dynamics are usually employed to describe predator-prey systems where the prey reproduces at a constant rate and the predator dies at a constant rate \citep{lotka:book:1925,volterra:JCIEM:1928}. 
This allows the population size to change.
The predator density is influenced by the abundance of prey 
and the prey density is influenced by the abundance of predators.
In the framework considered here, host-parasite interactions are abstracted to a level where they can be studied qualitatively in the same way as predator-prey dynamics. 
The complex underlying mechanisms are simplified substantially. 
For example, a linear birth rate of the host is considered, which would lead to unlimited exponential growth in the absence of parasites. 
Furthermore, since we do not use individual based models, but model population dynamics without including the individual level, it is not necessary to specify whether the parasites kill the host, reduce fecundity or slow down reproduction. 
These mechanisms are all reduced to few parameters: the interaction strengths. 
The interpretation of those is left open and can be different, depending on the species under consideration. 
The models thus focus on impacts on population dynamics. 

We assume $n$ different types of hosts and $n$ different types of parasites. 
The hosts have a constant birth-rate $b_h>0$ and a death rate that is determined by the interactions with the parasite.  
Conversely, we assume a constant parasite death-rate $d_p>0$, but a birth rate that depends on the interactions with the hosts.
With these assumptions the change of host ($h_i$) and parasite ($p_i$) abundance in time can be formulated as 
\begin{align}
&\dot{h}_i=h_i \left(f^H_i+b_h\right) &&\text{and} &&\dot{p}_i=p_i \left(f^P_i-d_p\right).
\end{align}
The fitness values $f^H_i = \left(M^H p \right)_i$ and $f^P_i = \left(M^P h \right)_i$ are defined as the interaction matrix multiplied with the abundances of the other species types.

Instead of immediately numerically exploring the dynamics for particular parameter sets, we first aim to
obtain some general insight. 
On the boundaries of the state space, we have one fixed point where all hosts and parasites
are extinct, $h_i = p_i =0$ for all $i$. 
In the absence of parasites, the host population will continue to increase in size, 
whereas a parasite population is not viable in the absence of hosts. 
In terms of co-existence, it is more interesting to consider potential interior fixed points.

The dynamics of the system depends crucially on the stability of the interior fixed point, which can be attracting, repelling or neutrally stable.
For the matching allele model, we have a fixed point where all
hosts and parasites have equal abundances, $h_i^*=d_p$ and $p_i^*=b_h$ for all $i$. 
In this case, the equations completely decouple and each host-parasite pair is independent of all others.
Thus, the fixed point is neutrally stable, as for the case of a single host and a single parasite.
This implies that a small perturbation from the fixed point does not lead back to it, neither does it increase the distance.  
For the cross-infection model, a host suffers from three parasite types and each parasite
type benefits from three host types. 
The internal fixed point is now $h_i^*=\frac{d_p}{3}$ and $p_i^*=\frac{b_h}{3}$. 
In terms of the cross infection models, it is substantially harder to prove the neutral stability, but in the Appendix~\ref{appendix:stability} we show that at least for $n\leq6$, the fixed point remains neutrally stable. 
For the generalised cross-infection model, we obtain $h_i^*=\frac{d_p}{\sum_{i=1}^n \alpha_i}$ and $p_i^*=\frac{b_h}{c \sum_{i=1}^n \alpha_i}$. 
For $n=3$, we can show that the fixed point remains neutrally stable if the interaction strength decreases with the distance between host and parasite type (see Appendix~\ref{appendix:stability}). \\
In addition, the symmetry of the system leads to a constant of motion. 
For all three models, the constant of motion is given by \citep{plank:JMP:1995,hofbauer:book:1998},
\begin{align}
H = \sum_{i=1}^{n} h_i - \sum_{i=1}^{n} h_i^* \log h_i + c \sum_{i=1}^{n} p_i - c \sum_{i=1}^{n} p_i^* \log p_i.
\end{align}
The existence of such a quantity arises from the symmetry of the system and implies that effectively, the system has one free variable less. 
\subsection{Constant population size}

As before, we assume $n$ different types of hosts and $n$ different types
of parasites. 
But now, $h_i$ denotes the relative abundance of host type $i$ and the relative abundance of parasite type $i$ is $p_i$
($i=1,2,\hdots,n$). 
With $h$ and $p$ we denote the vectors of the relative abundances. 
Thus, we have $\sum_{i=1}^n h_i=1$ as well as $\sum_{i=1}^n p_i=1$.
This means that the dynamics take place in a space of two dimensions less.
We assume that the relative abundances change according to the 
replicator dynamics \citep{hofbauer:book:1998},
\begin{align}
\dot{h}_i=h_i \left(f_i^H-\bar{f}^H\right) &&\text{and} && \dot{p}_i=p_i\left( f_i^P - \bar{f}^P \right)
\label{RD}
\end{align}
where $f_i^H = \left( M^H p\right)_i$ is the host fitness for type $i$ and $\bar{f}^H = h M^H p$ is the average fitness of the host population.
Similarly, $f_i^P = \left( M^P h \right)_i$ is the parasite fitness for type $i$ and $\bar{f}^P = p M^P h$ is the average fitness of the parasite population. 

For example, a system with two hosts and two parasites ($n=2$) where matching hosts and parasites have an influence of $\alpha_1=1$ and mismatching pairs exert a smaller fitness effect $\alpha_2=0.3$ with a twofold impact on the host $c=2$ would be a system of four differential equations,
\begin{align}
\dot{h}_1=&h_1\left( -2p_1-0.6p_2 - \left[h_1 \left( -2p_1-0.6p_2 \right) + h_2 \left( -0.6p_1-2p_2 \right) \right] \right)\\  
\dot{h}_2=&h_2\left( -0.6p_1-2p_2 - \left[ h_1 \left( -2p_1-0.6p_2 \right) + h_2 \left( -0.6p_1-2p_2 \right) \right] \right)\\
\dot{p}_1=&p_1\left( +1h_1+0.3h_2 - \left[ p_1 \left( +1h_1+0.3h_2 \right) + p_2 \left( +0.3h_1+1h_2 \right) \right] \right)\\
\dot{p}_1=&p_1\left( +0.3h_1+1h_2 - \left[ p_1 \left( +1h_1+0.3h_2 \right) + p_2 \left( +0.3h_1+1h_2 \right) \right] \right).
\end{align}

Again, this model can now be solved numerically to generate trajectories 
depending on the initial values of $p_1$ and $h_1$ (which determine $p_2=1-p_1$ and $h_2=1-h_1$). 
However, this approach would only lead to insights about particular parameter sets. 
Thus, here we take a different and -- in our opinion -- a more powerful approach
and look at general properties of the system. 

The replicator dynamics Eq.~\eqref{RD} has fixed points on the edge of the state space, e.g. $h_1=p_1=1$ and $h_i=p_i=0$ for $i>1$. 
However, these fixed points are saddles for a generic parameter choice, which means that a small perturbation from this point will eventually drive the system away.  
There is an additional fixed point where all types have equal abundance, $p_i^*=h_i^*=n^{-1}$ for all $i$. 
This arises from the symmetry of the interaction matrices we consider, but the fixed point can also be verified directly in Eq.~\eqref{RD}.

Again, for the matching allele model, the interior fixed point is neutrally stable for any number of types (see Appendix~\ref{appendix:stability}).
In the cross-infection model neutral stability is verified for $n\leq6$. 
Also for the generalised cross-infection model, an analysis is intricate, but for $n=3$, we again show neutral stability assuming that interaction strength decreases with distance.\\
There is a constant of motion, as recognized previously by \cite{hofbauer:JMB:1996}
\begin{align}
H &=\sum_{i=1}^n  \log{h_i} + c \sum_{i=1}^n \log{p_i}.
\end{align}
However, as we show numerically below, this does not imply any regularity of the dynamics. 

A model with $n=3$ types of hosts and parasites is thus not 6-dimensional, but 3-dimensional owing to the two constant population sizes and the constant of motion. 
However, even chaotic dynamics is possible in continuous-time systems of three dimensions.

\subsection{Irregular dynamics in the most simple model}  

While the general properties discussed above lead to a first insight, e.g. the fact that there is always an interior fixed point and that it is neutrally stable, they do not give insights beyond the fact that the dynamics is oscillatory near the fixed point. 
Close to the fixed point, one would expect regular oscillations, but it remains unclear what happens if we leave the vicinity of the fixed point. 

\begin{figure}[!htbp]
	\includegraphics[width=1\linewidth]{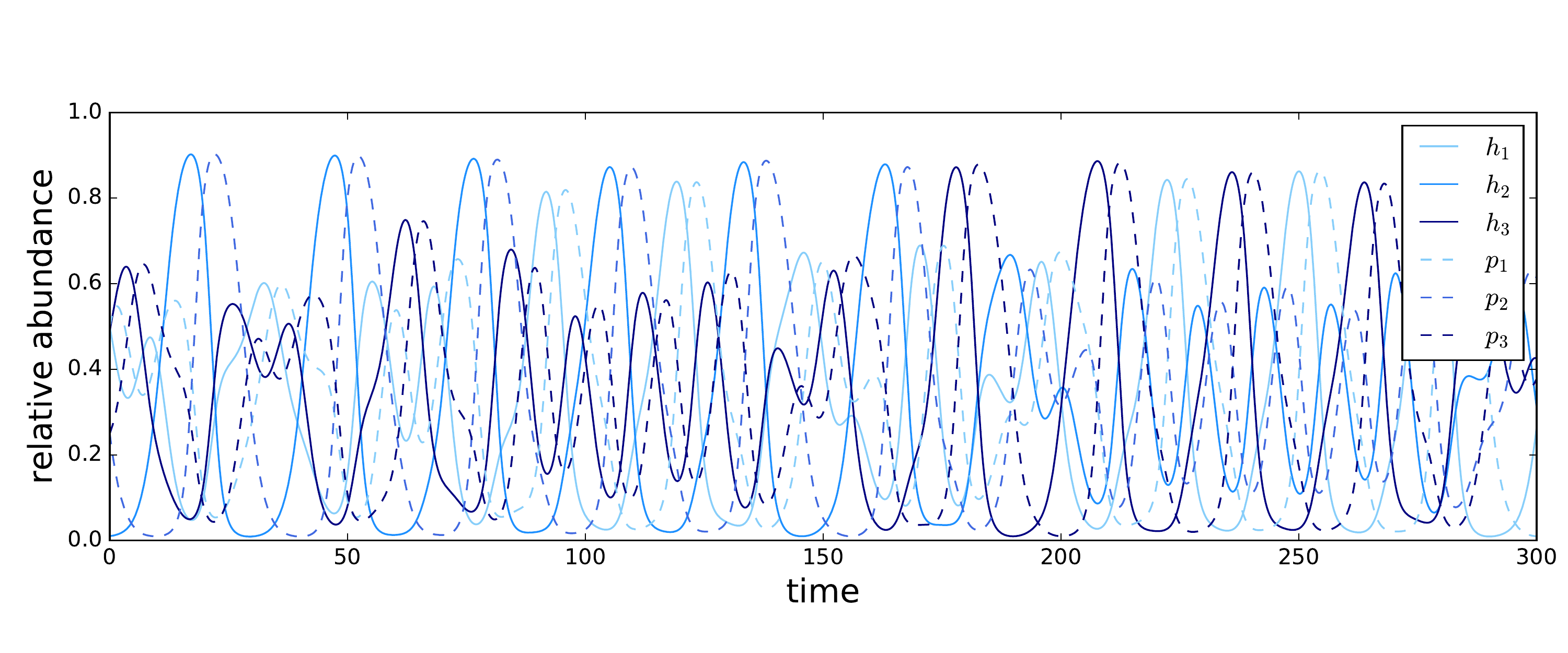}
	\caption{
	\textbf{Matching allele replicator dynamics with three types:}		The trajectories of all host and parasite types for a 3-type matching allele replicator dynamics system, Eqs.~\eqref{RD}, with initial conditions $h\left(0\right)=\left(0.5,0.01,0.49\right)^\top$ and $p\left(0\right)=\left(0.5,0.25,0.25\right)^\top$ oscillate, but not as regularly as often depicted in this kind of models. Numerical integration with \texttt{python}'s built-in \texttt{odeint} function.}
	\label{fig:trajectory}
\end{figure}

It turns out that in spite of the constants of motion and neutral stability, the trajectories of host and parasite abundances through time can become irregular and non-periodic. 
This can already be observed in a three type matching allele model with constant population size, interaction matrix $
M^P = 
\left( 
\begin{smallmatrix}
1 & 0 & 0\\
0 & 1 & 0\\
0 & 0 & 1
\end{smallmatrix}
\right)
$
, cf.\ Fig.~\ref{fig:trajectory}.  
This surprising result observed even in the simplest model we consider 
led us to examine this particular model more thoroughly.
For details on the numerical procedure see Appendix \ref{appendix:numerical}.

\begin{figure}[!htbp]
	\includegraphics[width=0.98\linewidth]{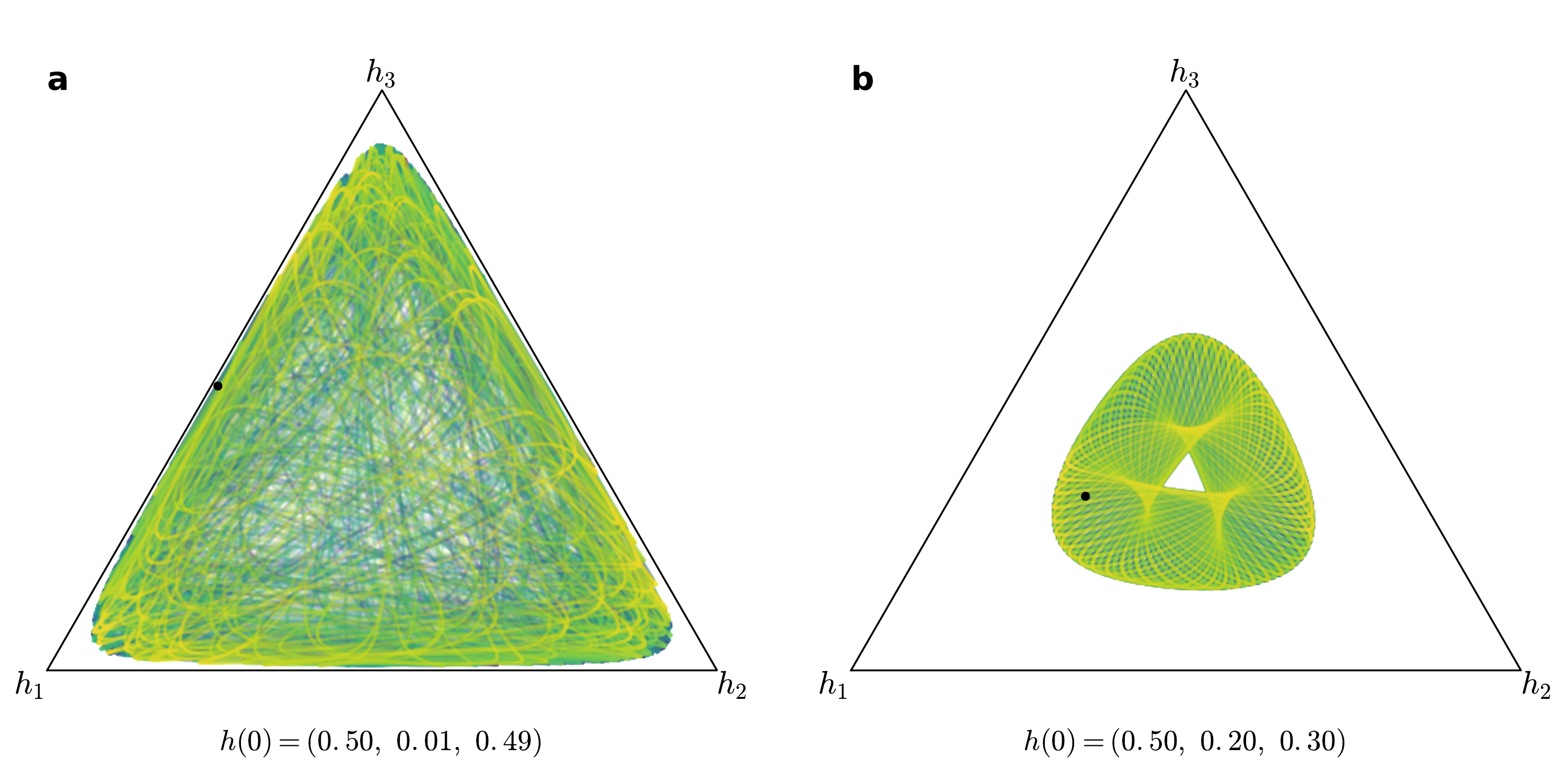}
	\caption{
	\textbf{Matching allele replicator dynamics with three types.}	
Host 3-simplex for a 3-type matching allele replicator dynamics system with initial conditions given by $h\left(0\right)$ and $p\left(0\right)=\left(0.5,0.25,0.25\right)^\top$ indicated as a black dot.
Panel (a) corresponds to the initial condition from Fig.~\ref{fig:trajectory}.  
Time is represented in the colour gradient going from dark to bright. For this initial condition close to the edge, irregular dynamics emerges.
(b) For initial conditions closer to the interior fixed point, the dynamics remains regular. 
Numerical integration was performed using  \texttt{python}'s built in \texttt{odeint}  function. 
Plotted until time $10000$ (a) and $5000$ (b).
}

	\label{fig:simplex}
\end{figure}

Because of the constant population size, a third type has a relative abundance determined by the abundance of the other two types. 
For each of the two species, the dimensions reduce from three to two.
It is therefore possible to show the dynamics for the three types of each species in a 3-simplex (Figure~\ref{fig:simplex}), where each vertex represents the sole existence of one type, the edges correspond to a coexistence of two types and the interior is a state where no type is extinct.
For balanced initial conditions close to the centre of the simplex, trajectories are confined to orbits around the interior fixed point.
For more extreme initial conditions, starting close to the edge of the simplex, this is no longer true. 
The trajectory is no longer limited to regular orbits, but nearly fills out the whole simplex, going from conditions close to extinction of one type (edges of simplex) to a near balance of all types (close to the interior fixed point).

\begin{figure}[!htbp]
	\includegraphics[width=1\linewidth]{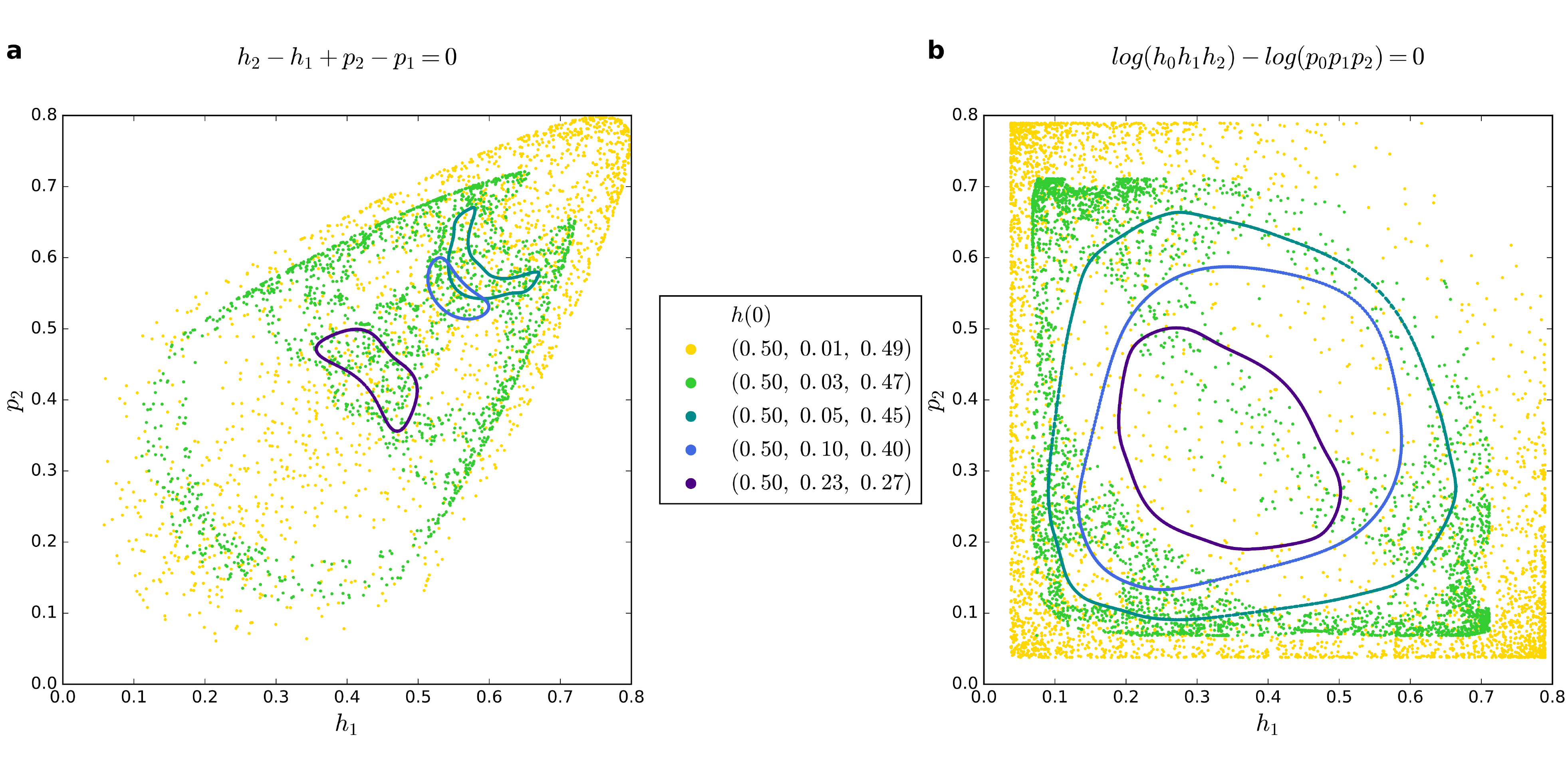}
	\caption{
	\textbf{Poincar\'{e} sections for a 3-type matching allele model with constant population size (replicator dynamics):} 
The Poincar\'e sections with two restrictions (a) (following \citet{sato:PNAS:2002}) and (b) are plotted
approximate such that all points in a small area around the two-dimensional plane are considered.
The horizontal and vertical axes are the host type 1, $h_1$ and parasite type 2, $p_2$ respectively. 
With initial conditions as stated in the legend and $p(0)=\left(0.5,0.25,0.25\right)^\top$. 
For initial conditions closer to the fixed point (lines, see also Fig.~\ref{fig:simplex}b), the trajectories show periodic behaviour in a higher dimension.
For extreme initial conditions, close to the edge of the state space (scattered points, see also Fig.~\ref{fig:simplex}a), the trajectories become chaotic and show a wide spread over the state space. Integration until time 100'000.}
	\label{fig:ps}
\end{figure}

To analyse the regularity of the dynamics further, we visualised trajectories of different initial conditions in Poincar\'e sections to check for chaotic behaviour \citep{strogatz:book:2000}.
Plotting Poincar\'e sections is a method to analyse dynamic properties of high dimensional systems.
This is implemented by plotting trajectories in a two-dimensional plane under certain restrictions (see Fig.~\ref{fig:ps}).  
Periodic trajectories pass through the section in a periodic way, drawing circles or other closed lines.
Chaotic trajectories have a much less ordered path and thus scatter over a larger part of the section.
\citet{sato:PNAS:2002} found chaotic behaviour and large positive Lyapunov exponents for several initial conditions in a two-person rock-paper-scissors learning game. 
This is formally closely related to a replicator dynamics host-parasite system with three types. 
We utilised this approach for our matching allele model and numerically evaluated several initial conditions. 
Quasiperiodic dynamics are visible as closed lines in Figure~\ref{fig:ps} for most initial conditions. 
For initial conditions close to the edge of the state space, however, the trajectories become visibly scattered.
These trajectories come very close to but do not re-visit all previous states.

\subsection{Similar dynamics with changing population size}
\begin{figure}[!htbp]
	\includegraphics[width=1\linewidth]{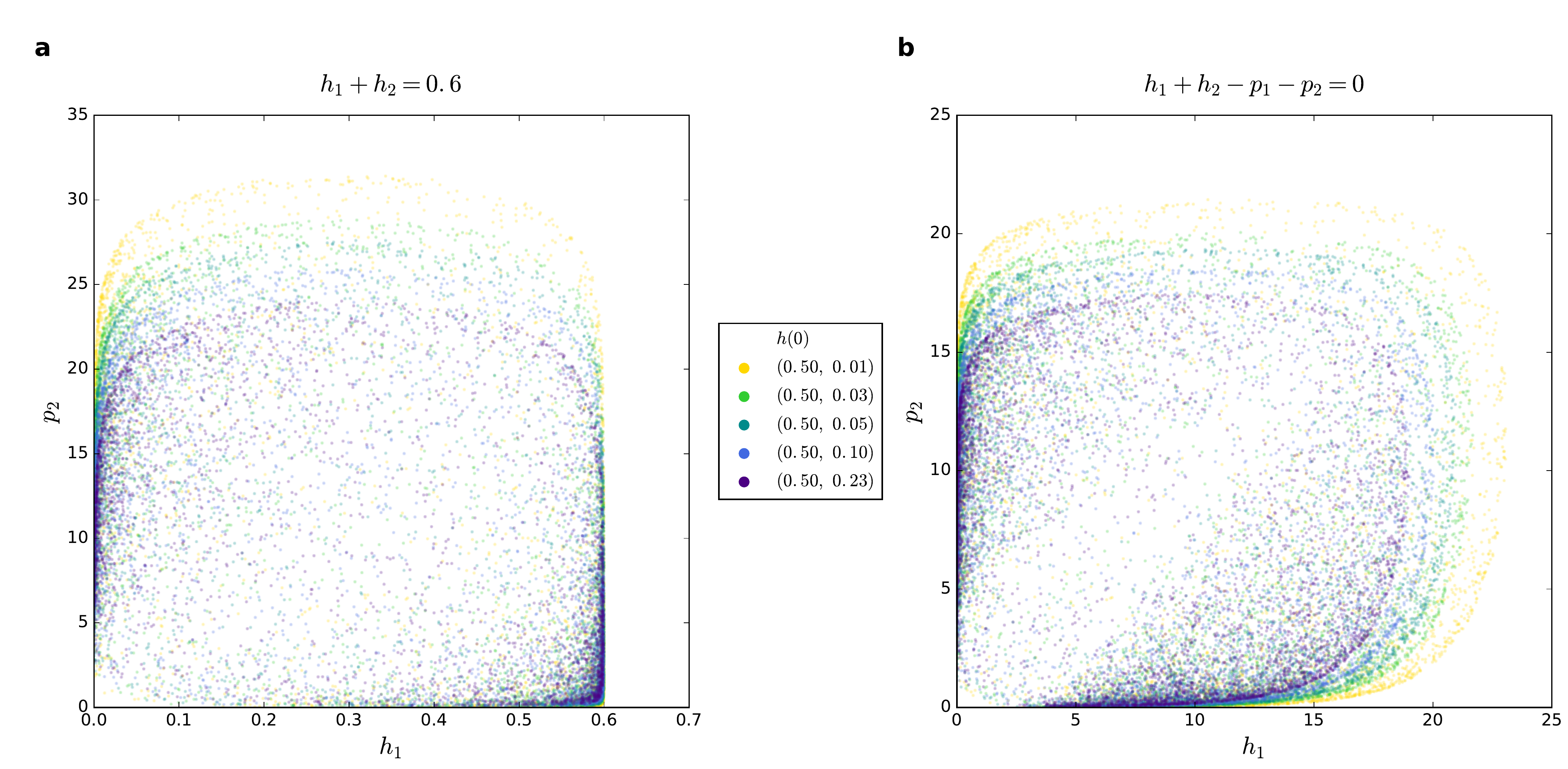}
	\caption{
		\textbf{Poincar\'{e} sections for a 2-type cross-infection model with changing population size (Lotka-Volterra):} 
		The Poincar\'e sections with two restrictions (a) and (b) are plotted.
		The horizontal and vertical axes are the host type 1, $h_1$ and parasite type 2, $p_2$ respectively. 
		With initial conditions as stated in the legend and $p(0)=\left(0.5,0.25\right)^\top$. 
		Parameters: birthrate $b_h = 5$, deathrate $d_p = 2.5$, infection by matching type $\alpha_1 = 1$, infection by cross-infecting type $\alpha_2 = 0.2$, no difference in impact on host or parasite $c=1$. Integration until time 10'000.}
	\label{fig:psLV}
\end{figure}

A similar dynamics is possible in models with changing population size.
Regarding the dimension of the system, a three-type model with constant population size as above compares to a two-type model with changing population size.
Whereas the matching allele model has coupled equations in the model with constant population size, the equations are completely independent of one another here.
The interaction matrix 
$
M^P = 
\left( 
	\begin{smallmatrix}
	1 & 0.2\\
	0.2 & 1
	\end{smallmatrix}
\right)
$
with matching pairs and a small influence of mismatching pairs allows for coupled equations.
With a constant of motion, the dimensions again reduce to three, still allowing chaotic dynamics.
The dynamics with changing population size (Fig~\ref{fig:psLV}) do not seem to depend on the distance of initial conditions to the fixed point in the same way as the model with constant population size.
We have set our focus on the model with constant population size, but hope to elucidate more on both models in a more thorough analysis.

\section{Discussion}

Originally, \citet{stenseth:Evolution:1984} as well as \citet{nordbotten:PNAS:2016} showed that only trophic  interactions (as opposed to mutualism or competition) promote Red Queen dynamics independent of abiotic factors.
This justifies our study of these dynamics in a simple framework without abiotic influence or other types of interactions than trophic.
We study host-parasite coevolution with three successively complicated interaction matrices in frameworks with more than two types where both constant and changing population size models can be justified \citep{macarthur:TPB:1970, papkou:Zoology:2016}.
In our work, we calculate specific outcomes with analytically derived statements and not only rely on numerical integration with fixed parameters and fixed initial conditions. 
For the same reason, we focus on a deterministic framework, allowing broad predictions. 
Eventually, one also has to include stochastic effects, which can have decisive impact on coevolutionary dynamics, in particular when the dynamics reaches the edges of the state space where extinction is likely \citep{gokhale:BMCEvolBio:2013}. Also to explore signatures of genomic selection, such as selective sweeps or balancing selection, this is necessary \citep{tellier:Evolution:2014}.
However, to obtain general results for the dynamics, we focus here on a deterministic framework.

Red Queen dynamics have been repeatedly reported to occur in models with two types, often because two alleles were in focus in the matching allele or gene-for-gene model \citep{schmid-hempel:AmNat:2002,frank:PRSB:1993,flor:AG:1956,agrawal:EER:2002,song:BMCEvolBio:2015}.
Models with multiple genotypes have been analysed, however often using more complex models.
Examples from observed biological systems clearly motivate the need for including this aspect more into theoretical studies \citep{carius:Evolution:2001,koskella:Evolution:2009,luijckx:Evolution:2014} and there is general interest in increasing also the number of species in the analysis \citep{liow:TEE:2011,dercole:PRSB:2010}.

\citet{rabajante:SciRep:2015} numerically investigated such host-parasite systems with multiple types:
The model used in their research is similar to the model by \citet{rosenzweig:AMNAT:1963} with Holling type II functional response and logistic growth for the hosts in absence of infection. 
The interactions are similar to a matching allele model, but the differential equations are coupled through a small background infectivity of all parasites. 
The study focuses on regular Red Queen oscillations, which can take place in higher dimensions, but the authors have shown that the parameter space in which this is possible reduces with the number of types. 
The approach differs substantially from this paper, since the investigations are  numerical and non-Red Queen dynamics are not investigated further. 

In our model, all internal fixed points are neither repelling nor attracting.
The presence of such neutrally stable fixed points and constants of motion may lead to the belief that Red Queen dynamics may generically be reflected by stable, regular orbits around the interior fixed point. These are also often shown to illustrate this kind of dynamics. 
A neutrally stable fixed point and the consequent concentric circles, spheres or higher dimensional oscillations around the point mean that the system is constantly changing, and yet, 
stationary in this change.  
Formulating constants of motions or Hamiltonians underlines this principle. 
However, the stability of a fixed point only holds locally and a constant of motion is a mathematical construct that may reflect a biological context.
In general, neutral stability can be connected with Red Queen cycling. 
To understand this kind of dynamics in more detail, we have focused on the simplest possible dynamics instead of attempting to construct a model for a concrete biological scenario. 
It is possible to consider more complex dynamics with stable or unstable interior fixed points or even limit cycles. However, our goal is to illustrate that even these simple models, which often form the basis for investigations of host parasite coevolution, can show a dynamics which is much richer than one would expect from verbal arguments or numerical considerations of such systems close to interior fixed points. 

Simple models built on differential equations have been famously known to show chaotic properties in the sense that close by starting conditions can lead to very diverse outcome, thus restricting the predictability of the dynamics to very short time horizons \citep{lorenz:JAS:1963,may:Nature:1976,hamilton:PNAS:1990,hassell:Nature:1991,schuster:book:1995,sato:PNAS:2002}. 
It thus comes as no surprise that a system of multiple interacting species can lead to chaos in some parts of the parameter space \citep{may:SIAMJAM:1975,smale:JMB:1976}.
Before going into the differences of previous studies to our model, we present the reader with some examples.

One example comes from the adaptive dynamics field, where evolutionary dynamics in a continuous trait space are analysed assuming that new mutants arise on a timescale slower than the underlying population dynamics.
Recently \citet{duarte:ActaB:2015} found chaos in the food chain model by \citet{dercole:PRSB:2010} (see also \citet{dercole:IJBC:2010}) with three traits, resulting in Red Queen oscillations in the trait dynamics.
This adaptive dynamics approach shows that chaotic dynamics (chaotic attractors) in the trait space are possible, while the underlying ecological dynamics (population dynamics, or allele frequency changes) are assumed to be more simple.
Our model in contrast, while omitting changes in phenotype space, allows for the presence of more than two distinct types at once and focuses on the resulting allele frequency change within this standing genetic variation.


\citet{sardanyes:CSF:2007} used a deterministic model based on ordinary differential equations. 
With logistic growth and a Holling type II functional response.
The model is based on the original model by \citet{rosenzweig:AMNAT:1963}, but with more types. 
The model is further refined by adding mutation and diffusion and a constant decay of species. 
The results show that chaotic behaviour is possible depending on the various parameters. 
The authors use the variance in amplitudes of local maxima as a measure for chaos (bifurcation diagrams). 

\citet{turchin:Ecology:2000} fitted a previously established predator-prey model to experimental data. 
The model includes sine functions to model seasonality in a framework of ordinary differential equations. 
The authors explain differences between northern more specialised predators, where chaotic dynamics are possible, and generalists in more moderate southern latitudes. They focus on finding parameters from data and then analyse those models. 
They calculate positive Lyapunov exponents for some models that describe dynamics in northern regions, but state that also other non-chaotic dynamics are possible.

Apart from differential equations, discrete maps are well known for chaos in lower dimensions.
A discrete time model with a matching alleles type of interaction \citep{seger:PTRSB:1988} also leads to apparently chaotic trajectories when at least three alleles (types) are considered.
 
Another way to increase dimensions and achieve chaotic dynamics is by combining predator-prey dynamics (with one type of each) with an epidemic model for the predator \citep{stiefs:MBEn:2009}. 
The authors showed that chaos can arise in general models (general Rosenzweig-MacArthur model linked with SIR model) and specified the results in a particular case with Holling type III functional response and an asymptotic incidence function.
  
These examples show that chaotic behaviour is not rare and should be expected in higher dimensional predator-prey or host-parasite models.
The difference between previous literature and our framework lies in the simplicity of the model of the matching alleles replicator dynamics. 
Another difference is that our system seems to depend on initial conditions, each leads to different dynamics without a common attracting point, limit cycle or chaotic attractor.
It is further interesting that in our case, chaos exists in parallel to a neutrally stable inner fixed point.
For multiple (three and more) types, we found that trajectories starting further away from the interior fixed point can show such chaotic behaviour.
Chaotic fluctuations of host and parasite abundances, therefore, become possible in parts of the parameter space. 
Envision a new type that is introduced to a system exhibiting typical Red Queen oscillations, e.g. via mutation or migration.
While the mutant appears at low frequencies, the system shifts to an edge in a higher dimension.
Our analysis predicts that this might often lead to chaotic dynamics rather than to stasis or the persistence of regular Red Queen oscillations.
The typical Red Queen dynamics is sometimes thought to consist of regular sinusoid-like oscillations of the frequencies of the different types within host and parasite populations with short periods and one or few amplitudes.
We are now facing highly irregular trajectories without periodic re-occurrence and different magnitudes of maxima in each cycle. 
Chaos, then, would be especially rampant in the presence of low levels of standing genetic variation, mutations and migration.
Moreover, it would in particular occur for very large populations, where the typical intuition of evolutionary biologists is to expect regular deterministic dynamics. 
With our model, we propose that in addition to the concepts of stasis or regular Red Queen cycling a third scenario - chaotic Red Queen dynamics - is possible and likely even in very simple frameworks.
We also suggest that these particular dynamics depend on the distance of initial conditions with respect to the inner neutrally stable fixed point.
A more detailed analysis of the chaos in this system is ongoing work and will be published in future with a more technical and thorough approach.

\section*{Acknowledgements}
We thank Hinrich Schulenburg for fruitful discussions and comments on the manuscript
and Heinz Georg Schuster, Christian Hilbe and Jens Christian Claussen for providing additional insights into nonlinear dynamics. 
Generous funding by the Max Planck Society is gratefully acknowledged. 
CSG acknowledges funding from the New Zealand Institute for Advanced Study and support from the Marsden Fund Council administered by the Royal Society of New Zealand.
The authors declare no conflict of interests.

\appendix
\section*{Appendix}
\section{Models}

\subsection{Changing population size: Lotka--Volterra dynamics}\label{LVallModels}

The specific models, defined by the matrices in Table 1 in the main text are now applied to Lotka--Volterra models. This leads to a differential equation for each $h_i$ and $p_i$, where $i=1,\ldots n$.\\

\textbf{Matching allele:}
We first focus on the matching allele model, where only matching types (pairs) influence each other.
Since $b_h$ and $d_p$ are constants, the differential equations are decoupled. 
We thus obtain $n$ independent systems of two differential equations each,
\begin{align} 
&\dot h_i =h_i\left(-p_i+b_h\right)  & &\dot p_i =p_i\left(h_i-d_p\right), 
\end{align}
where $i=1,\ldots,n$. 
This makes the Lotka--Volterra matching allele model a limiting case, with particularly simple dynamics.\\

\textbf{Cross-infection:}
The differential equations are now connected to each other by types $i\pm1$.
Because of the periodic boundary condition with $p_0=p_{n}$, $p_{n+1}=p_1$, $h_0=h_{n}$ and $h_{n+1}=h_1$,
it is possible to simplify the differential equations,
\begin{align}
&\dot h_i =h_i\left(-\left(p_{i-1}+p_i+p_{i+1}\right)+b_h\right) &&\dot p_i =p_i\left(\left(h_{i-1}+h_i+h_{i+1}\right)-d_p\right).
\end{align}

\textbf{Generalised cross-infection:}
Utilising the most general payoff matrices leads to these general differential equations
\begin{align}
&\dot h_i =h_i\left(\left(M^H p\right)_i + b_h\right)  & &\dot p_i =p_i\left(\left(M^P h\right)_i - d_p\right), 
\end{align}
where $p$ and $h$ are the vectors containing all population sizes $h_i$ and $p_i$ for $i=1,2,...,n$.
Depending on the entries of the matrix the equations can be decoupled (as with matching alleles) or coupled (as the cross--infection).

\subsection{Constant population size: Replicator dynamics} \label{appendix:RDmodels}

Next, we combine the interaction models from Table 1 with the replicator dynamics.\\

\textbf{Matching allele:}
Even though this model is based on interaction between matching types, owing to the constant population size there is an indirect effect of other hosts and parasites on one another. 
Biologically, this can for example reflect competition between hosts. 
If one host increases fast in numbers or when space, food or other resources are limited other hosts suffer from the increase of that specific type
and decrease in abundance. 
Applying the matching allele fitness effects to replicator dynamics leads to a set of coupled differential equations which describe the frequency change of host and parasite types, 
\begin{align}
&\dot h_i =h_i\left(-p_i+\sum_{k=1}^{n} h_k p_k\right)  & &\dot p_i =p_i\left(h_i-\sum_{k=1}^{n} h_k p_k\right). 
\end{align}

\textbf{Cross-infection:}
Infectiousness of neighbouring parasites leads to
\begin{align}
\dot h_i =&h_i\left(-\left(p_{i-1}+p_i+p_{i+1}\right)+\sum_{k=1}^{n} h_k \left(p_{k-1}+p_k+p_{k+1}\right)\right) \\
\dot p_i =&p_i\left(\left(h_{i-1}+h_i+h_{i+1}\right)-\sum_{k=1}^{n} p_k\left(h_{k-1}+h_k+h_{k+1}\right)\right),
\end{align}
with $p_0=p_{n}$, $p_{n+1}=p_1$, $h_0=h_{n}$ and $h_{n+1}=h_1$.

\textbf{Generalised cross-infection:}
The differential equations are now more complicated, so that it is best to present the general form,
\begin{align}
\dot h_i &=h_i\left(\left(M^H p\right)_i-h^T M^H p\right)  & \dot p_i &=p_i\left(\left(M^P h\right)_i - p^T M^P h\right).
\end{align}

\section{Stability} 
\label{appendix:stability}
A stability analysis of fixed points is conducted by calculating the eigenvalues of the Jacobian matrix at the interior fixed point.
The real part of the eigenvalues of this matrix gives insight into the stability of the point. 
If all are negative the fixed point is attractive, if at least one is positive it is a saddle (and repelling if all are positive) and if all are zero it is neutrally stable.  
These statements hold locally, which means close to the point of interest, since this is where the Jacobian is evaluated.
In the case of replicator dynamics it is possible to reduce the number of differential equations to $2\left(n-1\right)$ because of the normalisation $\sum_{i=1}^n h_i=\sum_{i=1}^n p_i =1$. 
The matrix now has full rank and the number of eigenvalues is always $2 \left(n-1\right)$.
The Jacobian is further explained below.
For deriving fixed points and stability for fixed $n$, \texttt{Mathematica} was used. 
Solving the differential equations $\dot{h_i}\left(h^*,p^*\right)=0$ and $\dot{p_i}\left(h^*,p^*\right)=0$ leads to some trivial fixed points where at least one host or one parasite type are extinct. 
We briefly discuss these fixed points for the matching allele model with replicator dynamics and three types below but focus on the non--trivial inner fixed point for all models.
\subsection*{Jacobian matrix} 
The Jacobian in the Lotka--Volterra case is in $\mathbb{R}^{2n \times 2n}$, where the number of eigenvalues is $2n$.
The lower dimensional Jacobian for replicator dynamics is
\begin{align}
J =
\left( \begin{array}{ccc;{2pt/2pt}ccc}
\frac{\partial \dot{h}_1}{\partial h_1} & \cdots & \frac{\partial \dot{h}_1}{\partial h_{n-1}} & \frac{\partial \dot{h}_1}{\partial p_1} & \cdots & \frac{\partial \dot{h}_1}{\partial p_{n-1}} \\
\vdots &  & \vdots & \vdots & & \vdots \\
\frac{\partial \dot{h}_{n-1}}{\partial h_1}  & \cdots & \frac{\partial \dot{h}_{n-1}}{\partial h_{n-1}}  & \frac{\partial \dot{h}_{n-1}}{\partial p_1}& \cdots & \frac{\partial \dot{h}_{n-1}}{\partial p_{n-1}}   
\vspace{3pt} \\ \hdashline[2pt/2pt]  \rule{0pt}{1\normalbaselineskip}
\frac{\partial \dot{p}_1}{\partial h_1} & \cdots & \frac{\partial \dot{p}_1}{\partial h_{n-1}} & \frac{\partial \dot{p}_1}{\partial p_1} & \cdots & \frac{\partial \dot{p}_1}{\partial p_{n-1}} \\
\vdots &  & \vdots & \vdots & & \vdots \\
\frac{\partial \dot{p}_{n-1}}{\partial h_1}  & \cdots & \frac{\partial \dot{p}_{n-1}}{\partial h_{n-1}}  & \frac{\partial \dot{p}_{n-1}}{\partial p_1}& \cdots & \frac{\partial \dot{p}_{n-1}}{\partial p_{n-1}}   \\
\end{array} \right)
\in \mathbb{R}^{2\left(n-1\right) \times 2\left(n-1\right)}.
\end{align}
For all subsequent calculations with constant population size (replicator dynamics) dimension were reduced by setting $h_n=1-\sum_{i=1}^{n-1} h_i$ and $p_n=1-\sum_{i=1}^{n-1} p_i$.
\subsection*{Lotka--Volterra: cross-infection} \label{jacLVCI}
For $n=4$, the eigenvalues are
\begin{align}
\lambda_1&=\pm \frac{i}{3} \sqrt{b_h d_p} \qquad \text{ each with multiplicity $3$,}\\
\lambda_2&=\pm i \sqrt{b_h d_p}  \qquad \text{each with multiplicity $1$.}\nonumber
\end{align}
For $n=5$, the eigenvalues are
\begin{align}
\lambda_1&=\pm \frac{i}{3\sqrt{2}} \sqrt{\left(3\pm \sqrt{5}\right)b_h d_p} \qquad \text{ each with multiplicity $2$,}\\
\lambda_2&=\pm i \sqrt{b_h d_p}  \qquad \text{each with multiplicity $1$.}\nonumber
\end{align}
For $n=6$, the eigenvalues are
\begin{align}
\lambda_1&=0 \qquad \text{with multiplicity $4$,}\\
\lambda_2&=\pm \frac{i}{3} \sqrt{b_h d_p}  \qquad \text{each with multiplicity $1$,}\nonumber\\
\lambda_3&=\pm \frac{2i}{3} \sqrt{b_h d_p}  \qquad \text{each with multiplicity $2$,}\nonumber\\
\lambda_4&=\pm i \sqrt{b_h d_p}  \qquad \text{each with multiplicity $1$.}\nonumber
\end{align}
In all cases analysed the interior fixed point is neutrally stable.

\subsection*{Lotka--Volterra: generalised cross-infection} \label{jacLVg}
The eigenvalues for $n=3$ are 
\begin{align}
\lambda_1=&\pm i \sqrt{b_h d_p} \qquad \text{ each with multiplicity $1$,}\\
\lambda_2=&\pm i \sqrt{b_h d_p}\frac{\sqrt{\alpha_1 \left(\alpha_1-\alpha_2\right)+\alpha_2 \left(\alpha_2-\alpha_3\right)-\alpha_3 \left(\alpha_1-\alpha_3\right)}}{\alpha_1+\alpha_2+\alpha_3} \\& \qquad \text{ each with multiplicity $2$.}\nonumber
\end{align}
Assuming that interactions with more distant hosts become weaker, $\alpha_1>\alpha_2>\alpha_3$, the term under the square root becomes positive and all eigenvalues have a real part zero. This implies neutral stability of the interior fixed point for $n=3$.

\subsection*{Replicator dynamics: Matching allele model} \label{jacRDMA}
After reducing dimensions the differential equations are defined for $i=1,2,...,n-1$ : 
\begin{align}
\dot h_i =h_i\left(-p_i+\sum_{k=1}^{n-1} h_k p_k + \left(1-\sum_{k=1}^{n-1} h_k\right) \left(1-\sum_{k=1}^{n-1} p_k\right)\right)
\end{align}
and 
\begin{align} 
\dot p_i =p_i\left(h_i-\sum_{k=1}^{n-1} h_k p_k - \left(1-\sum_{k=1}^{n-1} h_k\right) \left(1-\sum_{k=1}^{n-1} p_k\right)\right)
\end{align}
The Jacobian at the interior fixed point $h_i^*=p_i^*=\frac{1}{n}$ simplifies to
\begin{align}
J\left(h^*,p^*\right)=  \left( \begin{array}{ccc;{2pt/2pt}ccc}
&  &  & -\frac{1}{n} &  & \phantom{-}0 \\
& \text{\huge{0}} &  &  & \ddots &  \\
&  &   & \phantom{-}0 &  & -\frac{1}{n}  \vspace{3pt} \\ \hdashline[2pt/2pt]  \rule{0pt}{1\normalbaselineskip}
\frac{1}{n} &  & \phantom{-}0\phantom{-} & &  &  \\
& \ddots &  &  & \text{\huge{0}} &  \\
0  &  & \phantom{-}\frac{1}{n}\phantom{-}  &  &  &    \\
\end{array}\right) \in \mathbb{R}^{2\left(n-1\right)\times 2\left(n-1\right)}.
\end{align}
The eigenvalues of the Jacobian matrix are calculated via the determinant $det\left(J\left(h^*,p^*\right)-\lambda I_{2\left(n-1\right)}\right)$ which is not changed by adding multiples $\left( \frac{1}{\lambda n}\right)$ of the upper $n-1$ rows to the lower $n-1$. This leads to the following matrix
\begin{align}
\left( \begin{array}{ccc;{2pt/2pt}ccc}
-\lambda &  &  & -\frac{1}{n} &  & 0 \\
& \ddots &  &  & \ddots &  \\
&  & -\lambda \phantom{-} & 0 &  & -\frac{1}{n}  
\vspace{3pt} \\ \hdashline[2pt/2pt]  \rule{0pt}{1\normalbaselineskip}
&  &  & -\lambda-\frac{1}{\lambda n^2}&  & 0 \\
& \text{\huge{0}} &  &  & \ddots &  \\
&  &   & 0 &  & -\lambda-\frac{1}{\lambda n^2}   \\ 
\end{array}\right).
\end{align}
The determinant of this matrix is the product of the diagonal elements. The following equation determines the eigenvalues:
\begin{align}
0=& \left(-\lambda\right)^{n-1} \left(-\lambda -\frac{1}{\lambda n^2}\right)^{n-1}\\
=& \left(\lambda+i \frac{1}{n}\right)^{n-1} \left(\lambda-i \frac{1}{n}\right)^{n-1}\nonumber
\end{align}
The eigenvalues are $\lambda=\pm i \frac{1}{n}$ with multiplicity $n-1$. 
This means that the interior fixed point is neutrally stable and the oscillation frequency close to this point is $\frac{1}{2 \pi n}$. 
This implies that the period of the oscillation $2 \pi n$ depends on the number of types of host and parasite.
The oscillation with $n$ types has an oscillation period which is $n$ times longer than in a system with only one host and parasite type. 
In the case of replicator dynamics, this is based on the coupling through the average fitness $\bar{f}$. 
\subsection*{Replicator dynamics: cross-infection} \label{jacRDCI}
In the more complicated cross-infection model it is challenging to solve the problem for general $n$ which is why a stability analysis is shown for several fixed $n$.
For $n=4$, we find
\begin{align}
\lambda&=\pm  \frac{i}{4}  \qquad \text{each with multiplicity $3$}.
\end{align}
For $n=5$, we find
\begin{align}
\lambda &=\pm \frac{i}{5} \sqrt{\frac{1}{2}\left(3\pm \sqrt{5}\right)} \qquad \text{each with multiplicity $2$.}\nonumber
\end{align}
For $n=6$, we have
\begin{align}
\lambda_1&=0 \qquad \text{with multiplicity $4$,}\\
\lambda_2&=\pm \frac{i}{3}  \qquad \text{each with multiplicity $2$,}\nonumber\\
\lambda_3&=\pm \frac{i}{6}\qquad \text{each with multiplicity $1$.}\nonumber
\end{align}
The fixed point is neutrally stable in all cases.
\subsection*{Replicator dynamics: generalised cross-infection} \label{jacRDg}

In this generalised case the stability is analysed for $n=3$. The four eigenvalues depend on the payoffs:
\begin{align}
\lambda=\pm \frac{\sqrt{c}}{3} \sqrt{\pm \left(\alpha_2-\alpha_3\right)-\left(\alpha_1-\alpha_2\right)\left(\alpha_1-\alpha_3\right)}\label{eq:EVRD3}
\end{align}
The interior fixed point is neutrally stable if all eigenvalues are zero or without exception imaginary. 
This holds if the term under the square root is negative in all cases. 
This is the case if the second term is larger than the first, which implies $\alpha_1\gg\alpha_2>\alpha_3$. 
Thus, assuming that more distant hosts are less suitable for the parasite, the fixed point is neutrally stable.
If however there are two significant host types for each parasite and only one host is unsuitable $\alpha_1>\alpha_2\gg\alpha_3$ then two eigenvalues are real. In that case, we would have a saddle.
In other cases it depends on the value of the real part of the eigenvalues. When all eigenvalues are positive the fixed point is not stable (repelling) if all are negative then the fixed point is stable (attractive).

\subsection*{Trivial points}
The trivial fixed points, where all but one type of host or parasite is non-existent are saddles, which means they are stable in at least one direction but unstable in at least one direction.
For example, in the matching allele replicator dynamics model with constant population size and three types (Fig.\ref{fig:trajectory},\ref{fig:simplex},\ref{fig:ps}) the trivial fixed points are combinations of $h^* \in \{ (0,0,1), (0,1,0), (1,0,0)\}$ and $p^* \in \{ (0,0,1), (0,1,0), (1,0,0)\}$.
The Jacobian for this model is\\

\resizebox{\linewidth}{!}{%
$J(h_1,h_2,p_1,p_2)=\begin{pmatrix}
1 - 2 p_1 - p_2 + 2 h_1 (-1 + 2 p_1 + p_2) + h_2 (-1 + p_1 + 2 p_2) & h_1 (-1 + p_1 + 2 p_2) & h_1 (-2 + 2 h_1 + h_2) & h_1 (-1 + h_1 + 2 h_2)\\
h_2 (-1 + 2 p_1 + p_2) & h_1 (-1 + 2 p_1 + p_2) + (-1 + 2 h_2) (-1 + p_1 + 2 p_2) & h_2 (-1 + 2 h_1 + h_2) & h_2 (-2 + h_1 + 2 h_2)\\
-p_1 (-2 + 2 p_1 + p_2) & -p_1 (-1 + p_1 + 2 p_2) & -1 + h_2 + 2 p_1 + p_2 - 2 h_2 (p_1 + p_2) - h_1 (-2 + 4 p_1 + p_2) & -(-1 + h_1 + 2 h_2) p_1\\
-p_2 (-1 + 2 p_1 + p_2) & -p_2 (-2 + p_1 + 2 p_2) & -(-1 + 2 h_1 + h_2) p_2 & -1 + h_1 + 2 h_2 + p_1 - 2 h_1 p_1 - h_2 p_1 - 2 (-1 + h_1 + 2 h_2) p_2
\end{pmatrix}$}.\\

For example, inserting $(h^*,p^*)=(0,0,1,0,1,0)$ gives $J(h_1{=}0,h_2{=}0,p_1{=}0,p_2{=}1)=\left(\begin{smallmatrix}0&0&0&0\\0&-1&0&0\\0&0&0&0\\0&0&1&1\end{smallmatrix}\right)$ and therefore the eigenvalues $\lambda_1=-1$, $\lambda_2=1$, $\lambda_3=0$ and $\lambda_4=0$. 
Since both a positive and a negative eigenvalue is present the point $(h^*,p^*)=(0,0,1,0,1,0)$ is a saddle.\\

If we now allow arbitrary values for $p^*$ but keep $h^*=(0,0,1)$ the eigenvalues are $\lambda_1=1-p_2-2p_2$, $\lambda_2=1-2p_1-p_2$, $\lambda_3=-(1-p_1-p_2)$ and $\lambda_4=-(1-2p_1-2p_2)$ of which at least one is always negative and at least one is always positive when $p_1+p_2+p_3 = 1$ (constant population size), i.e. when $p_1 \in \left[0,1\right]$ and $p_2 \in \left[0,1-p_1\right]$.
The symmetry of the system ensures comparable results for all other configurations where in one of the species one type has taken over.\\

Relaxing the conditions further so that only one type of both species is extinct allows for complex eigenvalues.
The expressions have been numerically examined and so far only saddles have been found. 
\section{Numerical procedure} \label{appendix:numerical}

A stability analysis is a thorough way of determining the dynamical system's behaviour close to fixed points. 
To determine more properties of the dynamics, especially further away from fixed points, a simple approach is to numerically integrate the differential equations starting with a predefined initial condition at time zero.
Several techniques are available.
While the older, but most simple Euler and Runge-Kutta methods are easy to implement and to understand, modern solvers are often more reliable and accurate.
This is why we have chosen to use the more sophisticated solvers from the Fortran package for solving ordinary differential equations ODEPACK instead of our own implemented Runge-Kutta method of fourth order. 
All these solvers use an adaptive internal step size and return only those values determined by the sampling rate (defined by the user).
After obtaining qualitatively the same results with several of the selectable integration methods in \texttt{scipy.integrate.ode} we chose the more user-friendly \texttt{scipy.integrate.odeint}.
The latter uses the `lsoda' solver which selects dynamically between non-stiff and stiff methods.
The lowest initial condition is $0.01$ which is at the same time the lowest value in the numerical integrations for the replicator dynamics. 
Thus, numerical integration with values very close to zero is avoided.
For the Lotka-Volterra dynamics this may not be the case, but they are not the focus of this work.\\

Poincar\'e sections are lower dimensional planes (or manifolds) through a higher dimensional phase space. 
A plane is chosen by defining a condition which must be zero and then plotting all points occurring on that plane.
Because this plane has measure zero, a neighbourhood around the plane is chosen with a certain precision, which was set to $10^{-9}$ within the root detection algorithm. 
We chose two different manifolds to show that the choice does not influence the quality of the results.
The ODEPACK Fortran package contains an `lsoda' solver with a root detection facility `lsodar'. 
We have modified a python wrapper named \textit{odespy} by \citet{langtangen:software:2015} which we modified to do the following.
(1) Points in the phase space where the condition is met (roots are found) are saved, not only printed, for each integration.
And (2) the points are only saved when the trajectory passes through the plane from one direction, not from the other.\\

The code is published online at \url{https://github.com/HannaSchenk/RQchaos}.

\end{document}